\documentclass[10pt]{article}
\usepackage[explicit]{titlesec}
\usepackage{hyphenat}
\usepackage{ragged2e}
\RaggedRight

\usepackage{algorithmic}
\usepackage{multicol}

\usepackage{rotating}  
\usepackage{theorem}   
\usepackage{bm}  
\usepackage{amsmath,amsfonts,amssymb}
\usepackage{booktabs}  
\usepackage{pdfsync}   
\usepackage{graphicx}
\usepackage{latexsym}
\usepackage{amsmath}
\usepackage{amssymb}
\usepackage{fancyvrb}
\usepackage{subfigure}
\usepackage{colortbl}
\usepackage{enumerate}
\usepackage{pifont}
\usepackage{stmaryrd}
\usepackage{textcomp}
\usepackage{fncylab}
\usepackage{multirow}
\usepackage{paralist}
\usepackage{wrapfig}
\usepackage{longtable}
\usepackage[]{algorithm2e}
\usepackage[table]{xcolor}
\usepackage{lscape}
\usepackage{lipsum}

\usepackage{garamondx}
\usepackage[T1]{fontenc}
\usepackage{amsmath}
\usepackage{graphicx}
\usepackage{xcolor}

\usepackage{geometry}
\usepackage{fancyhdr}
\usepackage[labelfont={footnotesize,bf} , textfont=footnotesize]{caption}
\makeatletter
\renewcommand\tagform@[1]{\maketag@@@ {\ignorespaces {\footnotesize{\textbf{Equation}}} #1.\unskip \@@italiccorr }}
\makeatother
\titleformat{\section}
  {\normalfont}{\thesection}{1em}{\MakeUppercase{\textbf{#1}}}
\titlespacing\section{0pt}{0pt}{-10pt}
\titleformat{\subsection}
  {\normalfont}{\thesubsection}{1em}{\textit{#1}}
\titlespacing\subsection{0pt}{0pt}{-8pt}

\makeatletter
\newcommand\sixteen{\@setfontsize\sixteen{17pt}{6}}
\renewcommand{\maketitle}{\bgroup\setlength{\parindent}{0pt}
\begin{flushleft}
\sixteen\bfseries \@title
\medskip
\end{flushleft}
\textit{\@author}
\egroup}
\makeatother

\usepackage[sort&compress]{natbib}
\makeatletter
\renewcommand\@biblabel[1]{\textbf{#1.}\hfill}
\makeatother

\bibpunct{}{}{,~}{s}{,}{,}
\title{Towards Intelligent Risk-based Customer Segmentation in Banking}

\author{
Shahabodin Khadivi Zand*$^{a}$ \\ \medskip
$^{a}$Macquarie University, Sydney, Australia \\  \medskip
shahabodin.khadivi-zand@hdr.mq.edu.au
}

\begin{document}

\vspace*{.01 in}
\maketitle
\vspace{.12 in}

\section*{abstract}
Business Processes, i.e., a set of coordinated tasks and activities to achieve a business goal, and their continuous improvements are key to the operation of any organization. In banking, business processes are increasingly dynamic as various technologies have made dynamic processes more prevalent. For example, customer segmentation, i.e., the process of grouping related customers based on common activities and behaviors, could be a data-driven and knowledge-intensive process. In this paper, we present an intelligent data-driven pipeline composed of a set of processing elements to move customers' data from one system to another, transforming the data into the contextualized data and knowledge along the way. The goal is to present a novel intelligent customer segmentation process which automates the feature engineering, i.e., the process of using (banking) domain knowledge to extract features from raw data via data mining techniques, in the banking domain. We adopt a typical scenario for analyzing customer transaction records, to highlight how the presented approach can significantly improve the quality of risk-based customer segmentation in the absence of feature engineering.As result, our proposed method is able to achieve accuracy of \%91 compared to classical approaches in terms of detecting, identifying and classifying transaction to the right classification.

\section*{keywords}
Business Process Analytics; Data Curation; Customer Segmentation

\vspace{.12 in}


\section{introduction}
\label{chap:introduction}

This paper concentration is on understanding the necessity of technology development in the field of finance. One of the central focuses of the finance industry is to understand the risk appetite of the customer~\cite{Fin-cris-2008-3}. Based on the risk factors defined in the field of money lending, it is essential to ensure the customers are reliable and trustworthy for funding. Banks provide intelligent business analysis on their customers and a background check to ensure their investment is being used for the proficient reasons\cite{HistoryBank}. It is crucial to analyze customer's financial behaviour and patterns for gaining a better knowledge of their risk behaviour~\cite{HistoryBank2}. 

Research has been conducted on the current impacts of technology, process analytic, and business process on banking. Moreover, an intelligent data-driven pipeline has been introduced that can extract critical information from a customer's data and enrich them with available data domains. It would provide an insight into their background and allows the financial institution to connect them correctly to their risk metrics~\cite{HistoryBank2}. Later, a measure of success of the proposed method is measured to ensure the proposed system is effective and efficient.

\subsection{Motivation}

Money lending has been one of the leading and long in the tooth products of banking, and it provides an opportunity for people and organizations to achieve their dreams by making them possible. Customers of the bank consist of various type of people with different preferences and agenda. By the development of technology, there have been many changes in methods of banking during the 21st century. Cash and coins that have been known as the most valuable currencies, nowadays are being replaced by blockchain money such as Bitcoin~\cite{Bitcoin}. The course of banking has been altered due to recent developments, and the risk of fraud and financial crimes have tremendously increased during recently. Financial Technology also known as Fin-Tech, seeks to provide and implement most recent technologies in the banking system. Applying techniques such as personalization of products based on customer preference has been one of the long dreams of industries. A vast number of companies are investing in their business process and business intelligence to get more insight into their business and customers~\cite{Fin-Tech1,Fin-Tech2}.

These recent changes in the field of Fin-Tech have brought the attention of many companies and governments to develop several systems that can understand people's behaviour and predict their actions. It has motivated the author of this paper to begin a scientific investigation in the current issues of the financial sector and use recent intelligent feature engineering in the field of risk-based customer segmentation. 

\subsection{Problem Statement}

The quality of the services any organization provides largely depends on the quality of their Business Processes, i.e., a set of coordinated tasks and activities to achieve a business goal~\cite{ProcessAnalytics,ProcessAtlas}. 
Accordingly, business processes and their continuous improvements are key to the operation of any organization~\cite{ProcessAnalytics,BPM}. 
In the edge of Big Data - i.e., a massive number of small data islands from personal, shared, social, and business data - the business world is getting increasingly dynamic and many processes are becoming data-driven and knowledge-intensive~\cite{DataSynapse,CDCR}.
For example, in banking, processes are getting increasingly dynamic and knowledge workers are heavily involved in the execution of those processes and choosing the best next steps.

For example, Risk-Based Customer Segmentation, i.e., the process of identifying the risks by grouping related customers based on common activities and behaviors, requires knowledge workers such as compliance officers to understand the banking big data and quickly adapt to new regulatory requirements (and changes in how bad actors conduct financial crime) through risk-Based transaction monitoring and analytical segmentation.
This process involves risk-based transaction monitoring by combining customer/bank accounts with similar properties (as well as transaction behavior) to facilitate formulating risk signals based on various classes of customers in banking systems.
In such an ad-hoc process, developing an analytical segmentation modeling requires identifying the characteristics that helps understanding the bank’s customers data (such as transaction type, volume, amount, descriptions) and contextualize them with external knowledge sources.
This process is quite knowledge intensive and requires involving domain experts in the segmentation process to: perform data quality checks on the customer attributes, and to perform an exploratory analysis of the results to make sure outliers are removed.

\subsection{Contribution}

This paper contributes by presenting an intelligent data-driven pipeline composed of a set of processing elements to move customers' data from one system to another, transforming the data into the contextualized data and knowledge along the way. 
By utilizing an intelligent risk-based customer segmentation process, system can automates the feature engineering, i.e., the process of using (banking) domain knowledge to extract features from raw data via data mining (to automate turning the raw data into contextualized data and knowledge) and crowdsourcing~\cite{CrowdCorrect,MoCrowd,MoCrowd2} (to mimic the knowledge of a banking domain expert) techniques, in the banking domain.
We adopt a typical scenario for analyzing customer transaction records, to highlight how the presented approach can significantly improve the quality of customer segmentation in the absence of feature engineering.

\subsection{Summary and Outline}

In this section we have provided a general description of the problem that this research attempts to solve. We explained the motivations that have led the author to identify the research problem in the financial sector, as well as the contribution of this paper. The rest of this dissertation is organized as follows:

\textbf{section 2} discusses an overview of the banking industry and how it was established in history. Different types of financial institutions and their services are briefly explained to the reader to familiarize them with the concept of banking. Essential concepts and needs of banking are presented to bring readers attention to the impact of banking on daily life. Then, the effects of technology development on the banking services and organizations are discussed to highlight the benefits and drawbacks of it. Moreover, a general understanding of business intelligence and business processes in banking are introduced by the author to underline the importance of process analytic and customer segmentation in the finance sector.
\textbf{Section 3} presents the methods and algorithm taken to achieve the aims of the paper. This section will also provide a detail description of the proposed approach and how the implementation of it has progressed. 
\textbf{Section 4} briefs the reader about the type of dataset that was used to evaluate the proposed system and the system requirements to execute it. Then, a motivating scenario is described as an example of a real-world problem that could be solved by the proposed intelligent data-driven pipeline. 
\textbf{Section 5} concludes the methodology and results of the proposed method and what has been achieved by the end of the paper. Also, it provides plans and recommendation that could enhance the performance of the system and simplify more challenges in the business sector. 

\section{Background and State-of-the-Art}

This section will provide an overview of the banking industry and how it has control over the economy of the world. Moreover, different types of financial institutions will be introduced to the readers and familiarize them with their operations. Then, technology-led banking and how information technology (IT) has impacted the banking process, from 1950 to the present time, will be investigated in details. Next, products and demands of customers on the banking sector will be discussed to provide a detailed map of customer segmentation. 

\subsection{Banking Industry Overview}
It is not clear to determine the exact date of banking, considering it started before the writing age. Many theories have concluded that banking was initiated from ancient empires to keep the treasure safe during wars. Temples and other houses of worships were initially considered as safe, and a place to get a loan.  These holy places used coins or seeds as currency, since storing them where much simpler compared to others. Moreover, as civilization had grown, the banking process got developed too. The history of modern banking can be traced back to the late 15th century in Italy, where the word \textit{Banca} was used to refer to people who conducted different businesses on benches \cite{HistoryBank}. The oldest operating bank is B\textit{anca Monte dei Paschi di Siena} located in the heart of Italy. This bank was opened in 1472 and is known as the beginning of modern banking. Since 1400, the banks started new financial services such as money transfer, currency exchange and money lending. Thus far, financial services have been one of the top industries in the world as they maintain and control the economy of the world. This sector consists of a broad range of industries like commercial banking, insurance companies and investment banking~\cite{HistoryBank2}. Money lending or loan product is known as on the most popular services of banks with a long history. 

In August 2007, the most disastrous worldwide financial crisis started happening around the world. It immensely affected the banks and financial institution and moving them to the edge of bankruptcy. Governments had to step in and support them to avoid bankruptcy. Millions of people had lost their job and houses due to crisis. The global financial crisis was caused by several reasons and had impacted all the stock markets around the world~\cite{Fin-cris-2008-1}. The top three reasons were incorrect pricing of risk, comfortable credit conditions and wrong banking model. Incorrect pricing of risk refers to having a wrong scorecard that is not able to calculate the minimum amount of money that the individual could pay back risk-free. Risk has a direct effect on the interest rates and fees; therefore, having a wrong risk assessment could cause underestimating or overestimating of an individual's assets. The second important cause was reducing the interest rate of loans to almost \%50 lower from 2000 to 2003~\cite{Fin-cris-2008-2}. This action resulted in creating a housing bubble on mortgages, as well as forces governments to borrow from aboard, which could result in currency drops. Moreover, deploying a wrong servicing model into a banking system could have a significant effect on lending to medium or large businesses~\cite{Fin-cris-2008-3}. 

\subsubsection{Types of Financial Services and Institutions}

Currently, financial services and intuitions are categorized into several types and provide a variety of products to customers. The most highlighted type of financial service providers are mentioned below~\cite{FinancialIns}:

\begin{itemize}
    \item Commercial Bank is known as the most secure places for deposit accounts and lending to the public. These banks are mostly owned by governments.
    \item Credit Unions are similar to banks, however, their customers are a specific range of people such as military personal or farmers of a specific region. Shareholder of a Credit Union is consisted of its members and would operate in the interest of their members. 
    \item Insurance companies would assist the customer in determining the risk of loss of an asset and provide financial support in times of need.  
    \item Broker firm is responsible to provide a platform to securely sell and purchase stocks among investors. 
\end{itemize}

These institutions provide a variety of services to their customers, these services have different risk factors and interest rates~\cite{Financialservices}:
\begin{itemize}
    \item Personal loan refers to borrowing an amount of money from a bank and paid back over a certain period with constant interest. Dependable on the risk of the borrower, the loan would be secured or unsecured. 
    \item Business loan is mostly similar to a personal loan with the exception that borrower must be a registered firm or business.  
    \item Mortgage is a special type of loan that is provided specifically for home and land purchase. this loan is mostly long term and has a significant low-interest rate.
    \item SACC loan is a specific type of short term loan that is below a certain threshold and repayments are paid in a year. SACC loans are considered as high-risk loans and have a high-interest rate. 
    \item Debtor finance provides a customer cash flow of business in return of their slow-paying invoices, this would allow the customer to have a higher finance capital. 
\end{itemize}

\begin{figure}[t!]
    \centering
    \includegraphics[width=1\textwidth]{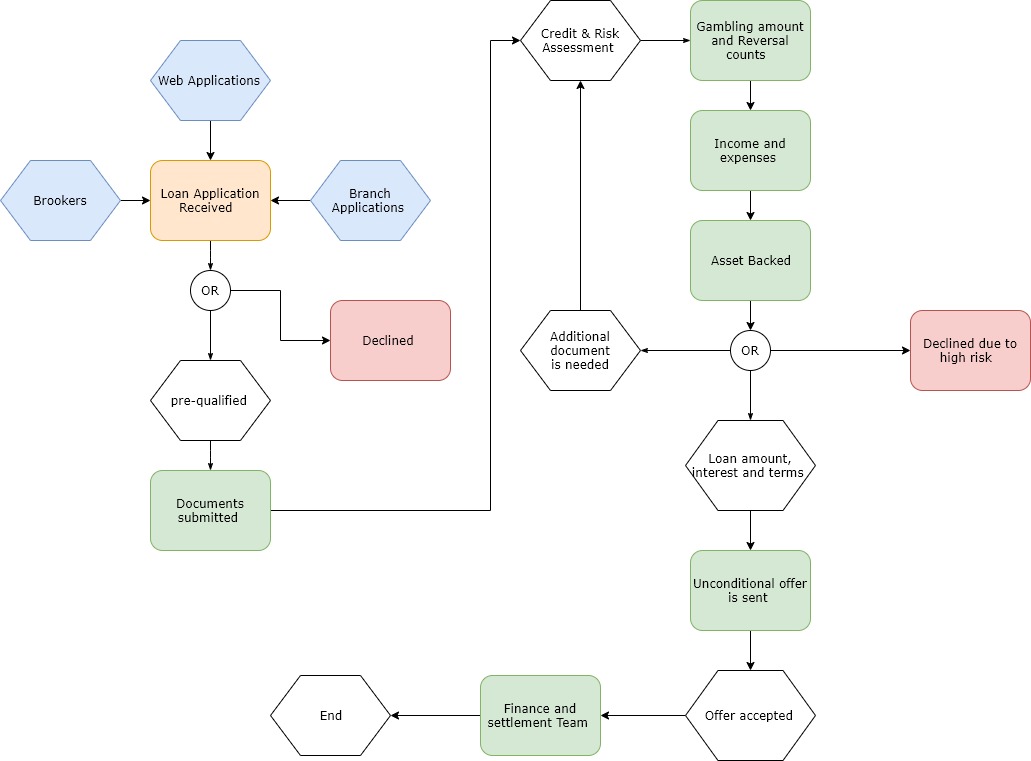}
    \caption{A general flowchart of a loan process in financial sector.}
    \label{fig:Bank_diagram}
\end{figure}

In general, providing a personal loan is still the first demand of customers and therefore the primary product of any financial institution. Figure \ref{fig:Bank_diagram} illustrates the general process of banking process for a loan application, and the steps that it has to take through the process. To lodge any bank application, it is essential to provide documents to the firm. If and only if all the documents are received, the financial firm could start analysing the customer's behaviour~\cite{p2vec,vahidBehaviour} and financial status. Based on the customer profile, the bank would provide a product that could suit both parties of the contract.

\subsubsection{Evaluation of Balanced Scorecard}

In every banking product, the most important factor in gaining a better performance evaluation of the process. To do so, a system called Balanced Scorecard has been introduced that looks at an organization from four perspectives which are customer, financial, internal business process and learning growth~\cite{scorecard1,benatallah2016business1,benatallah2016business2}. 
The financial perspective of scorecard focuses on the revenue and profits of the business. More precisely, it investigates the expense and income of an organization concerning their performance and budget. Every financial institution has a scorecard to evaluate their customer's behaviour and analyse their capacity for profitability~\cite{scorecard1,scorecard3}. 

One of the main elements of every scorecard would be the credit score of the customer. Credit score refers to a numerical score that indicates the history of the customer and its reputation in being a borrower. This score is ranged between 0 to 1200 that indicates the level of trustworthiness and assist the financial institutions to be a responsible lender~\cite{scorecard1}. Credit score could be used as one of the features of the scorecard, as it provides insight into the customer's behaviour in the past and would help to understand the operational risk~\cite{scorecard4}.  

\subsubsection{The Need For Banking}
Having a bank account would mostly have a positive impact on the daily life of people, by providing factors such as safety, credit transfer, saving access, and global reach. Another factor that emphasizes the need for banking would be investments of banks in the economy~\cite{needbank1}. Banks would mostly invest their profit in the margin of the market and propellant of the economy. Having access to the large scale of branches provides opportunities for investors to seek and discover potentially profitable businesses in rural areas. it would result in the development and growth of the rural area~\cite{needbank2}. 

\subsection{Technology-led Banking}

By the beginning of computing age in 1950, one of the applications of first-generation computers were to perform mathematical calculations. It was at this period that different technologies such as Automatic Teller Machine (ATM), Funds Transfer Pricing (FTP) and electronic cash counter were introduced~\cite{ATM1}. By 1980, the era of digital globalization was getting introduced, which highlighted the process of having a global network and sharing information across the country's borders. In terms of banking, the process of digital globalization used technology to provide services such as internet banking, phone banking, payment gateways and mobile banking~\cite{techbank2}. This era has been continuously growing ever since until today, where concepts of blockchain technologies and cordless transactions are being funded by that idea. In this day and age, technology is moving the banking sector into real time decision making and better customer segmentation~\cite{techbank1}. 

This section will investigate how the financial industry has been influenced by the technology timeline and provides a better understanding of the impacts of technology in the banking process. Therefore, the banking process that is most impacted by information technology are the transformation of money-dealing, credit scoring of individuals, internationalization of banking and digital money.

\subsubsection{Transformation of Money-dealing}

Before 1960, most of the banks had many branches all over the world, and it allowed them to interact with their customers. The typical process of banking would have required the customer to be present at the branch for any process. Teller would be in charge of assisting the customer's enquiry during the daytime of weekdays, which limited the operating time of banks. Therefore, two different groups of scientist and engineers introduced Automated Teller Machine (ATM) to the world~\cite{ATM2}. Being the first inventor of ATM has sparked a heated debate among scientists since two separate engineers introduced two different types of ATM at the same time in two different countries. The machines are not similar in terms of design, but they provide the same functionality. The first ATM was installed in 1967 at Barclays' Enfield Town branch. Later in 1970, the usage of PIN as an authorization method provided a secure transaction with the machines. PIN had a huge impact in terms of money dealing transactions since it eased up the process of security and verification of account holder. It would ask the account holder to set a four to six digit number as a personal identification number (PIN)~\cite{ATM3}. With the invention of ATM, many bank branches have closed over the years, or they have transformed to e-branches. E-branch refers to a bank location that is filled with several ATM and is operating full-time. 



\subsubsection{Credit Scoring}

The concept of credit scoring refers to the worthiness of an individual in terms of applying for a loan. Credit score was introduced in 1990 in the USA, with respect to the mortgage prices. The Australian government issued a set of protocols known as the Banking Code of Practice that requires financial institutions to be a responsible lender. This act requires banks and lenders to analyse and estimate the correct credit score, so the customer would be able to pay back without substantial hardship. Collection of the required information is costly and time-consuming, which give rise to a platform for technology to take action~\cite{CreditScoring1,CreditScoring2}. 
 
Nowadays, credit scoring is widely used by finance companies to calculate the score of the applicant based on the performance of their previous loans and predict their capability to pay back. For instance, if an applicant has a certain amount of income, the lender should take the income into a statement and calculate the total amount that the applicant can pay over the requested term. Several attributes have a role in determining the credit score of the individual, which would differ from one institution to another~\cite{CreditScoring3}. Each financial institution has its aggregate risk measurements that could be based on different probability theory and statistics. These mathematical frameworks are used to determine the dependencies between random variables that could affect the applicant. In Banking, random variables could be substitute with facts and information of the customer and define a model for the bank to obtain the risk of the product. 

To create and develop a scorecard, a minimum number of 30,000 loan applications are required to provide a statically confidence model~\cite{CreditScoring4}. This was the first step to involve technology in credit scoring, it would allow the firm to have access to the information and create databases. Over the years, this process has prompted banks to be memory centres and database of society and has risen issues such as data privacy and securities. Then, by performing the mathematical theories and statistical concepts, firms can provide a numerical assessment of the individual using technology. These developments point to current days when credit scoring and credit decision making are affected by new technologies such as artificial intelligence and machine learning~\cite{CreditScoring5,intelKG}. It is predicted that by 2030 all the credit scoring of people would be done by new technologies and AI~\cite{CreditScoring6}. In section 2.3, the author will review other applications of AI in the banking process and how it could affect customer segmentation. 

\subsubsection{Internationalization of Banking}

By 1990, the invention of technology has concentrated the focus of firms on banking rather than the bank. It was during this time that Bill Gates stated that "Banking is necessary, banks are not". The motivation supporting these statements was the rise of the internet and data. Data refers to a set of raw information and facts~\cite{InBank1}. By having a further understanding and analysis of data, an analyst can observe relationships and patterns between information~\cite{iStory}. Before 1980, most Banks used to create, collect and store their customer information and portfolio manually on paper. With the invention of relational databases and Structured Query Language, banks started transforming their manual documents into digital databases. This action allowed easier access to data and stock heap of data. 

Over the past decade, there has been an increasing interest in financial data such as bank statements and portfolio. Having an enormous volume of data that is so large to process has emerged introducing the concept of big data to the financial sector. Each banking transaction whether internal or external of an individual creates data that could be harnessed by different technologies. Big data is defined in terms of three main vectors which are variety, velocity and volume. By applying these vectors to the financial industry, it can be observed that having various data types such as structured and unstructured~\cite{InBank2,rajabi2016interlinking}. In terms of velocity, depending on the size of the firm and the number of customers the speed of new data is soaring. Moreover, by increasing the availability and accessibility of the internet and mobile banking, the volume of data has become incredibly high too~\cite{InBank3}. 

The arguments presented above shows the impact of information technology and the internet on the financial industry, how establishing a worldwide connection could help to increase the vectors of big data. Internet banking was initiated by Stanford Credit Union which was the first created online banking system in 1994. Seven years later, major banks of US provide online banking services to almost 19 million households across the country~\cite{InBank1}. Nowadays, online banking and mobile banking is a standard practice that could be done gratis and effectively. 

Another impact of technology could be observed on the sales team of the financial industry, as they try to use phone technology as technical support and sale of the product. Using technologies such as voice recognition are used to provide a secure platform for customers to gain access to their accounts. Other systems such as voice recording are being used to provide better training to customer service representatives at call centres~\cite{InBank4}. These recordings help them to prepare better for the unexpected circumstances that could happen. Moreover, these recording could also be used management and marketing team of a firm to produce better policies or marketing strategies for its customers. 

The great relative importance of internationalization of banking is the ability to transfer instantly across the globe without any fee. Gateway payment technology such as PayPal, XE Money Transfer and eWay provide features for customers and act as an electronic wallet~\cite{InBank1}. Most of these gateways are available on smart devices and provide a high-security platform for their customers. Another aim of these gateways is to provide currency exchange at low rates, so the customers can purchase or transfer to any country. 

\subsubsection{Digital Money}

One of the most recent impacts of technology on banking was the invention of Bitcoin. Bitcoin is a digital currency that is currently being used to purchase goods. To better understand Bitcoin, one must investigate the concept of cryptocurrency which are valued based on blockchain technology. Block-chain technology refers to having a set of blocks connected chronologically. Each block contains verified transaction data which are stored inside it. Then, each block will be added to the blockchain and will be publicly accessible~\cite{digmo1,digmo2}. 

Bitcoin is the first application of blockchain technology that was aimed to reduce the government's control over the cash flow of transactions and speed up the transactions. Therefore, Bitcoin is not legally accepted by large companies or governments, cause they might use it for money laundering. For instance, the Iranian government has announced a plan to shift some of its currencies into bitcoin, so it would undermine the US's sanctions. The blockchain is the technology responsible to hold the transactions of Bitcoin and makes it adaptable for any company that might require the information. 

Furthermore, more business is starting to accept bitcoin as a payment, which could be a threat to the financial sector specifically banks. This technology could be altering the traditional way of banking and change customer behaviours. As a result, Bitcoin is pushing the bank industry to become more digitized and provide real-time services~\cite{digmo3}. Another issue that is being arisen by having digital money is the ownership of the central bank. If banks decide to leave the traditional way and operate as digital banks, the ownership of central banks will be given to the public or the bank. These impacts of digital money are currently being investigated by researchers and governments to provide a confident and secure platform to customers~\cite{digmo2}. 

\subsection{Business Intelligence in Banking}

Nowadays, business intelligence has a magnificent role in banking due to having a variety of data across different data sources. Banking data is gained from ATM transactions to online shopping, and having access to these many information does provide the bank with a massive quantity of data. On the other hand, quality of data is ineffective. As discussed in section 2.2.3, the concept of big data is applicable to financial data, since it has the velocity, variety and volume. Business intelligence is the transformation of big data into more insightful data to help the organizations to take action~\cite{BI1}. The financial industry could use business intelligence to get a deeper understanding of customer segmentation. Customer segmentation consists of customer behaviour, geographic customer segment, customer values and Psychographic Segmentation. 

As a result of applying business intelligence to the finance industry, the operation efficiency of the company could be improved. Seeing the major expense of financial firms relates to their operational expenses, improving operational efficiency could decrease this factor. The main benefit of applying business intelligence to a firm is to decrease the operational risk in terms of tracking employee behaviour and analyzing credit portfolios of the firm~\cite{BI2}. Moreover, optimizing the marketing strategies would also be a benefit of applying BI to the organization data. This would provide a clear target for the marketing team in terms of understanding customer behaviour. For instance, knowing the need of customer could help the product team to modify or provide a custom setting for the product pending on customer. By the same token, personalization of product could also help the financial firms to gain an upper hand against their direct and indirect competitors. Direct competitor refers to a firm that provides the same product or services as your company, and indirect refers to a company that provide not similar products and services but targets the same market and customers~\cite{BI3}. 

\subsubsection{Behavioural Segmentation}

Behavioural Segmentation defined the ways and circumstances that the consumer would decide to use the product. Patterns such as affordability, attitude, and their response to a product, would determine the likeliness of recommending or re-ordering that products~\cite{bhS1}. For instance, millennials prefer to watch a movie over reading a book, in contrast, generation X prefers otherwise. This type of knowledge would help the financial firms to develop more targeted approaches to their customers. Targeted approaches would aid the firm in terms of personalization, predictive, prioritization and performance of segments over time. One of the main application of behavioural segmentation is purchasing behaviour, which indicates how a customer would approach the decision and what complexities would be in the purchasing process~\cite{bhS1,bhS3}.  

The complexity of purchase is considered as one of the main deal-breaker in the financial industry. According to the department of finance of Australia, a mortgage loan process would usually take 5 to 8 weeks for approval, and it requires many documents to be submitted. This issue has provided the private sector with an edge over governmental organizations due to having less number of clients. The private sector has simplified the loan process in the aim of being competitive to governmental organizations~\cite{bhS4}. Identically, having a variety of products could also complex the purchase order by providing confusion over which product would serve them best. As a consequence, the customer would be in doubt whether they are making the right choice or are they being manipulated into purchasing without knowing. 

\subsubsection{Geographic Customer Segment}
Another type of customer segmentation is geographic segmentation which refers to grouping the customer based on their geographical location. This act would allow the business companies to gain a better understanding of the demands of the area and neighbourhood. For instance, bushfires have caused hundreds of Australians to lose their farms and equipment, having this knowledge allows the loan provider to target a specific demand for that location by providing equipment loan product. In terms of marketing, knowing a specific geographic customer demand would all the financial firms to understand selling and advertisement strategies to expand their business~\cite{ge1}.  

\subsubsection{Customer Values}
Generally, the revenue of a business is defined as the income of that business from operational activities. In banking, a revenue has mainly consisted of the loan's interest, account fees, and credit cards. Forecasting an accurate revenue would result in planning the supply, cash flow management and better credit management~\cite{fore1}. Therefore, the customer values can be defined as the number of products and services purchased by the customer, and the performance of the firm will be determined. One of the main challenges of financial firms is to forecast the correct revenue for their business~\cite{fore2,fore3}. The Sydney Morning Herald highlights the most compelling evidence about a commercial company experienced a share drop of 27.7 per cent as the impact of having the wrong revenue forecast. Consequently, many stakeholders were a detriment, and the company had to reduce the operating cost by 7 million dollars. It has been reported that the actual revenue is 8\% lower than the forecast, due to having wrong credit grading~\cite{smh}. 

Furthermore, several techniques and models are used to calculate an accurate forecast for banks such as time-series technique, regression techniques(machine learning) and artificial intelligence techniques. According to the scientific work of Makridakia and Hiblon, the newer machine learning models are not performing ahead of traditional methods in time-series techniques. This result shows the simplicity of mathematical calculations provides better accuracy compared to complex ones~\cite{makridakis2018statistical}. In terms of AI, Jian and Qingyuan have conducted applying a backpropagation neural network to forecast the product sale of a company. Their result shows a good agreement that manually calculated values, and less than 4\% error is an acceptable result~\cite{jian2020market}. The new approaches are the foundation of future forecasting techniques and help the financial industry to understand their customer values by predicting the product and services they will purchase. This would allow the company to forecast revenue and provide the best values to its customer. 

\subsubsection{Psychographic Segmentation}
One of the rare analysis on segmentation belongs to psychographic segmentation, also known as the attitude and lifestyle of the customer. It divides and sorts the customers into contrasting groups, based on the based on their lifestyle, opinion and interests. This segmentation provides an insight into how customers behave based on their needs and respond to marketing strategies. Execution of this segmentation requires research on market and demand of the customers. By enabling artificial intelligence, a system would be able to analyse and understand people's reaction toward a particular marketing strategy. For instance, when a business starts having a giveaway offer for its customers, data extracted from the sale database could help to analyse the market's behaviour toward this strategy~\cite{psy1}. 

\subsection{Business Processes in Banking}

In the modern world, every business is aiming to achieve the most efficient and effective method of execution of their business plan. Having success in implementation would highly be beneficial in terms of arranging each individual's activities and ensuring the efficient use of resources. Automation is one of the main elements to help a business succeed. But, to apply automation methods, one must simply understand the and analyse the business processes related data. Data could be accessed based on the business model and implemented systems. 

\subsubsection{Process Analytic in Money Lending}

\begin{figure}[h]
    \centering
    \includegraphics[width=1\textwidth]{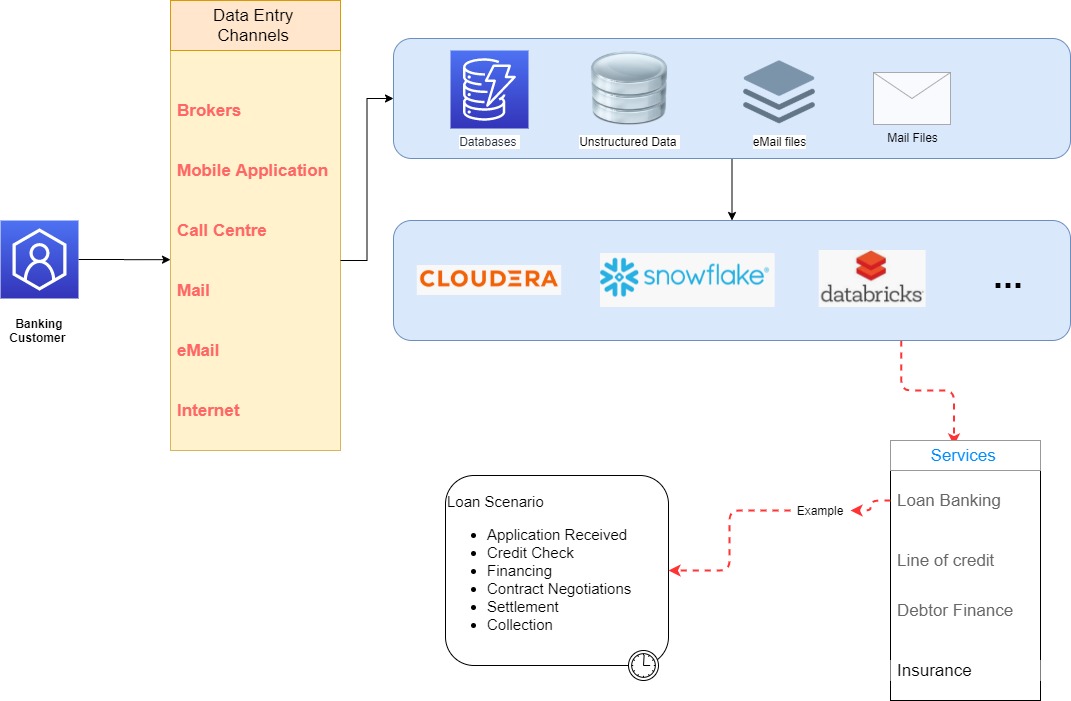}
    \caption{An example of a business process in finacial sector.}
    \label{fig:BP_in_banking}
\end{figure}

The key to improving business success in the financial industry is the automation of money lending process. Automation could be implemented through the whole business process from when an application is lodged all the way to settlement stage. In terms of risk analysis, a variety of data are required from services such as knowledge-based, cloud-based and domain experts to help the automation process~\cite{PA1,DBLP:journals/fgcs/SerhaniKSNBB20}. Figure~\ref{fig:BP_in_banking} shows an example of a business process in terms of banking. 

By having a closer look at the figure above, it illustrates a scenario where a customer is applying for a small business loan to increase their capital funding. Consider Yas(a customer) that owns a local bakery store and is interested in applying for a small business loan with an amount of less than a 10,000AUD. The process of money lending has various steps; the first step would be finding a suitable lender. Several ways, such as using a broker, cloud engines(e.g., Google) and financial accountants could be used to allocate the right financial institution. Next, she should need to apply and submit her documents through one of the channels such as mail, web services or broker. Meanwhile, she could start looking for some reviews on social platforms and recognised financial services like Google review, Trustpilot and Mortgage \& Finance Association of Australia (MFAA) to gain more understanding on the benefits and drawbacks of different institutions and their services. In terms of risk analysis, Yas's bank statement will go through a separate process to analyse and calculate her average monthly revenue (AMR) and average monthly expenses(AME). AMR and AME would provide insight and analysis of Yas's business behaviour. 

By having a closer look at Yas's application, it can be observed that several data sources are required to update the board of directors regarding the business process. Ability to answer questions regarding the status of each loan application demands a large amount of data processing which is a challenge. Each application should be processed and analysed to be able to answer fundamental questions such as ``What is this money used for? Is the applicant related to any politics? Is there a conflict with a competitor? Is the business legit and recognised? Is it related to any other application? Can it be a money laundering? Is this money being used for terrorism''. Answering these question highlights the value of process analytic in terms of banking and money lending. 

\subsubsection{Banking Processes}

According to Van Der Alast, a business process is defined as multiple tasks being processed automatically or manually to achieve a business goal~\cite{van2003business}. A business process could be classified into two categories which are public and private. Public business process refers to a process that is being shared among partners such as brokers and other financial institutes. In contrast, a private business process is confidential and can not be shared with anyone outside the enterprise. Most of the business process in terms of the application process are considered as public in banking; however, if the business process relates to scorecard and credit score, they are most probably confidential~\cite{van2004business}. Every organisation uses a set of methods and techniques to govern, develop and present their business process called business process management(BPM)~\cite{PA2}. There is much software such as Mavim, Decisions and Asana designed to manage operational business process. In terms of banking, BPM software helps to govern and control the lending process's life cycle. Each BPM has four stages defined to propose a process life cycle, which is design, configuration, enactment and diagnosis. 

In the context of the finance sector, business process management helps the automation, risk framework and understanding the customer. One of the most efficient ways to improve the business process of banking is implementing artificial intelligence to overtake manual process. Having an automated BPM would allow the financial organisation to gain a better understanding of the internal process and monitoring every stage of it. In addition, by providing an overview perspective on the process, it could simplify the process for customers and optimise the process based on customer demands. Another benefit of using BPM software in baking is making the decision making quicker and more efficient in terms of risk framework~\cite{van2004business}. Having a defined detailed process would accelerate the portfolio reporting and risk analysis. 

\subsubsection{Process Analytic in Banking}

To understand the performance of a business process and ways to improve it, the notion of process analytic is introduced~\cite{ProcessAnalytics}. It would assist the stakeholders in getting an insight on questions such as "Decline reasons, What happened to application \#453? Where is the application lodged? Who was the last person to update the application?". The most highlighted challenge to answer the following question is having access to the right data sources. Having high volume, different type and structure, trustworthiness and velocity of data, implies using the notion of Big Data in process analytic. It indicates capturing, organizing and processing of data are required to perform a successful analytic~\cite{PA2}. 

In terms of capturing data, several methods could be used to address creating rich metadata. Each technique could be utilized to request queries and store the data in a particular section of servers. One of the most recent methods is the usage of data management in cloud services. In this case, third party companies such as Microsoft Azure and Amazon are offering services to scale and distribute the dataset in their servers across the planet. Another method would be utilizing APIs to access data from service providers. In Australia, financial intuitions use Equifax database to obtain credit reports and credit score of their customers. Moreover, enterprises could utilize Dataspaces to organize and manage data based on a specific requirement~\cite{PA1}. For example, a bank might request a background check on their customer based on a keyword in a database. 

Organizing the captured data could also be one of the challenges in process analytic in the banking sector. Being able to arrange the captured data in a way to assist the computer processing step is the aim of enterprises. Looking at the captured data in terms of big data platforms by having a high volume, suggests utilizing techniques that could simplify the processing of it. Firstly, all captured data should be cleaned and arranged based on the systems and services. It includes removing errors from the logs, mapping, integration and filtering based on the requests of the financial firm. Then, integrating and graphing the essential entities and their relationship. In the context of banking, integrating the gambling behaviour of the customer could affect the risk framework of a bank and the application's outcome. A technique that could accelerate the processing time of a massive amount of data is Map Reduce. By using this method, data will be distributed among different data nodes, and the address of each input will be stored in name nodes~\cite{DataSynapse}. 

\subsubsection{Data-Driven and Knowledge-Intensive Processes in Banking}

\subsubsection{Data-Driven Processes in Banking}
The current Banking system is quite outdated, and the opportunities in the edge of Big Data (i.e., a massive number of small data islands from personal, shared, social, and business data) the entire financial ecosystem is undergoing radical change~\cite{DataSynapse,iProcess}.
Understanding the big banking data will provide personalized, trusted customer experiences, as well as meet risk and compliance mandates.
One of the main challenges in this domain is to transform banking data into actionable insights.
To achieve this goal it will be vital to prepare and curate the raw banking data for analytic. Data
curation has been defined as the active and on-going management of data through its lifecycle
of interest and usefulness~\cite{freitas2016big}. 
Data curation includes all the processes needed for principled and controlled data creation, maintenance, and management, together with the capacity to add value to data~\cite{arocena2016benchmarking}. 
The aim of this paper aim is at banking data creation and value generation, rather than
maintenance and management of this banking data over time. More specifically, the focus is on curation
tasks that transform raw banking data (e.g., a transaction record) into contextualized data and
knowledge include extracting, enriching, linking, and annotating banking data.

\subsubsection{Knowledge-Intensive Processes in Banking}
Knowledge-intensive processes contain a set of coordinated tasks and activities, controlled by knowledge workers to achieve a business objective or goal~\cite{iRecruit,iSheets}.
These processes are primarily semi-structured, since they often require the ongoing intervention of skilled and knowledgeable workers, and heavily reliant on professional knowledge. For these reasons, it is considered that human knowledge workers are responsible to drive the process, which cannot otherwise be automated as in workflow systems~\cite{ProcessAtlas,iCOP}.
In banking sector, many processes are knowledge-intensive and almost always involve the collection and presentation of a
diverse set of artifacts and capturing the human activities around those artifacts. This also emphasizes the artifact-centric nature of such processes.

Such knowledge-intensive banking processes, controlled by knowledge workers who have the experience,
understanding, information, and skills. Therefore, it will be vital to mimic the knowledge of a banking domain expert and build a domain specific Knowledge Base to facilitate the automation of data-driven and knowledge-intensive banking processes.
In this paper, the added value of our approach, compare to the state of the art in risk-based customer segmentation, is to focus on understanding the big data in banking sector, provide automated techniques curate the raw data, annotate the curated banking data with the knowledge of the banking experts, and relate this curate-annotated banking data to the process analysis.
To achieve this goal, this research presents an intelligent data-driven pipeline composed of a set of processing elements to move customers' data from one system to another, transforming the data into the contextualized data and knowledge along the way. We present an intelligent risk-based customer segmentation process which automates the feature engineering, i.e., the process of using banking domain knowledge to extract features from raw data via data mining (to automate turning the raw data into contextualized data and knowledge) and crowdsourcing (to mimic the knowledge of a banking domain expert) techniques, in the banking domain. 

\subsection{Summary}
To summarise, this section discusses the state of the art on how banks were founded and how their process is impacted by technology through time. Several known types of financial organizations and their products are highlighted and explained. Then, the method of evaluation of customer credibility is disclosed and how a bank could determine the trustworthiness of an applicant. Moreover, most known impacts of technology on the banking process are described, and the processes affected by the development of technology. Implications of technology on the transformation of money-dealing, credit scoring, internationalization and digital money were debated comprehensively. Furthermore, business intelligence in banking was discussed in terms of banking and how banks could understand their customers. Segmentation classes such as behavioural, geographic, values and psychographic were discussed in details. Next, an overall explanation of the business process is provided to ensure readers are familiar with the concept of BP and business process management(BPM). 

\section{Methodology}
\begin{figure}[t]
    \centering
    \includegraphics[width=1\textwidth]{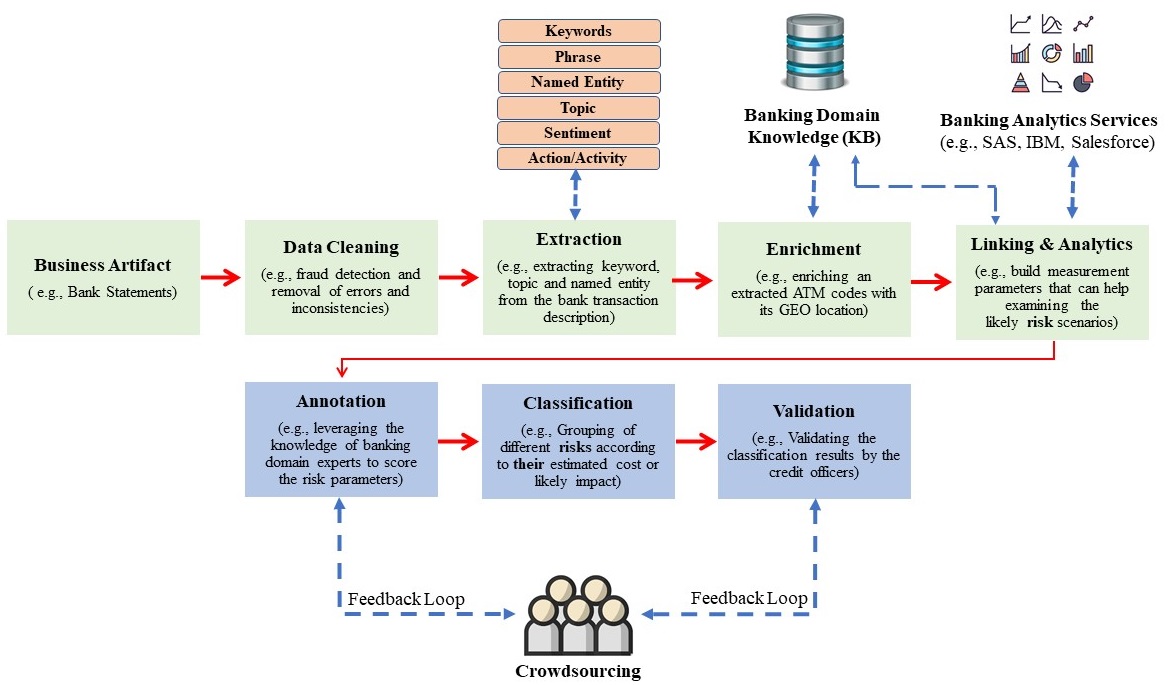}
    \caption{The proposed intelligent data-driven pipeline to automate risk-based customer segmentation process in banking.}
    \label{fig:framework}
\end{figure}

This section provides detailed planning of the proposed method and algorithm to achieve the aims of the project. By going through this section, the reader would gain a better understanding of the process of the proposed method and details of the system. It also presents an intelligent data-driven pipeline composed of a set of processing elements to move customers' data from one system to another, transforming the data into the contextualized data and knowledge along the way, and use the knowledge of domain experts to build a domain-specific Knowledge Base (KB) in banking.
%
%
In this section, we present an intelligent data-driven pipeline composed of a set of processing elements to move customers' data from one system to another, transforming the data into the contextualized data and knowledge along the way. 
The goal is to present an intelligent risk-based customer segmentation process which automates the feature engineering, i.e., the process of using (banking) domain knowledge to extract features from raw data via data mining (to automate turning the raw data into contextualized data and knowledge) and crowdsourcing (to mimic the knowledge of a banking domain expert) techniques, in the banking domain. Figure~\ref{fig:framework} illustrates the proposed framework. In the following an explanation on the main components of the intelligent data-driven pipeline is provided to the reader.

\subsection{Banking Data Curation}

The Big Data problem can be seen  as a massive number of small data islands from personal, shared, social, and business data. In this context, data curation (i.e., the process of transforming the raw data into contextualized data and knowledge~\cite{CoreDB,curateAPI,CoreKG}) play a vital role.
In a banking scenario, the business artifacts such as bank transactions, customer records, loan applications and more; are considered as first-class citizens and thus an abstraction to focus the process on core informational entities that are most significant~\cite{ProcessAnalytics}.
Therefore, such an artifact-centric approach may have hidden insight and knowledge that are defined in the context of process-related artifacts, and can be discovered by applying data curation techniques to such business artifacts.

For example, a bank transaction may have information about the type of the device (e.g., mobile or PC) used to submit the transaction, location information (e.g., IP address and GEO information), time of the transaction and the description of the transaction.
As the first stage in our pipeline (Figure~\ref{fig:framework}), we focus on curation tasks (include extracting and enriching) that transform raw banking data (e.g., bank transaction) into contextualized data and knowledge.

\subsubsection{Extraction Phase}

We present a business artifact as a data object that exists separately, has a unique identity and can be described with an attribute vector $[Item_{id}, Item_{type}, Item_{schema}]$ where,
$Item_{id}$ is a mandatory attribute whose value represents the unique identity of the business artifact;
$Item_{type}$ is a mandatory attribute whose value indicate the type of the business artifact; and
$Item_{schema}$ is a mandatory attribute whose value represents the schema of the $Item$.
Examples of a business artifact can be a bank transaction.
During the extraction phase, we extract various features including~\cite{DataSynapse}: 
(i)~Schema features: for example, a bank transaction may include information about time, location, source, and description;
(ii)~Lexical features, e.g., words or vocabulary of a language such as Keyword, Topic, Phrase, Abbreviation, and Special codes (e.g., ATM code or bank code).
(iii)~NLP features, e.g., entities that can be extracted by the analysis and synpaper of Natural Language such as Part-Of-Speech (e.g., Verb, Noun, etc), Named Entity Type (e.g., Person, Organization, Product, etc), and Named Entity (e.g., a bank name or a customer name).
(iv)~Time features, e.g., mentions of time in the schema of the item (e.g., transaction time or a date mentioned in the description of the transaction).
(v)~Location features, e.g., mentions of locations in the schema of the item.

\begin{figure}
    \centering
    \includegraphics[width=0.65\textwidth]{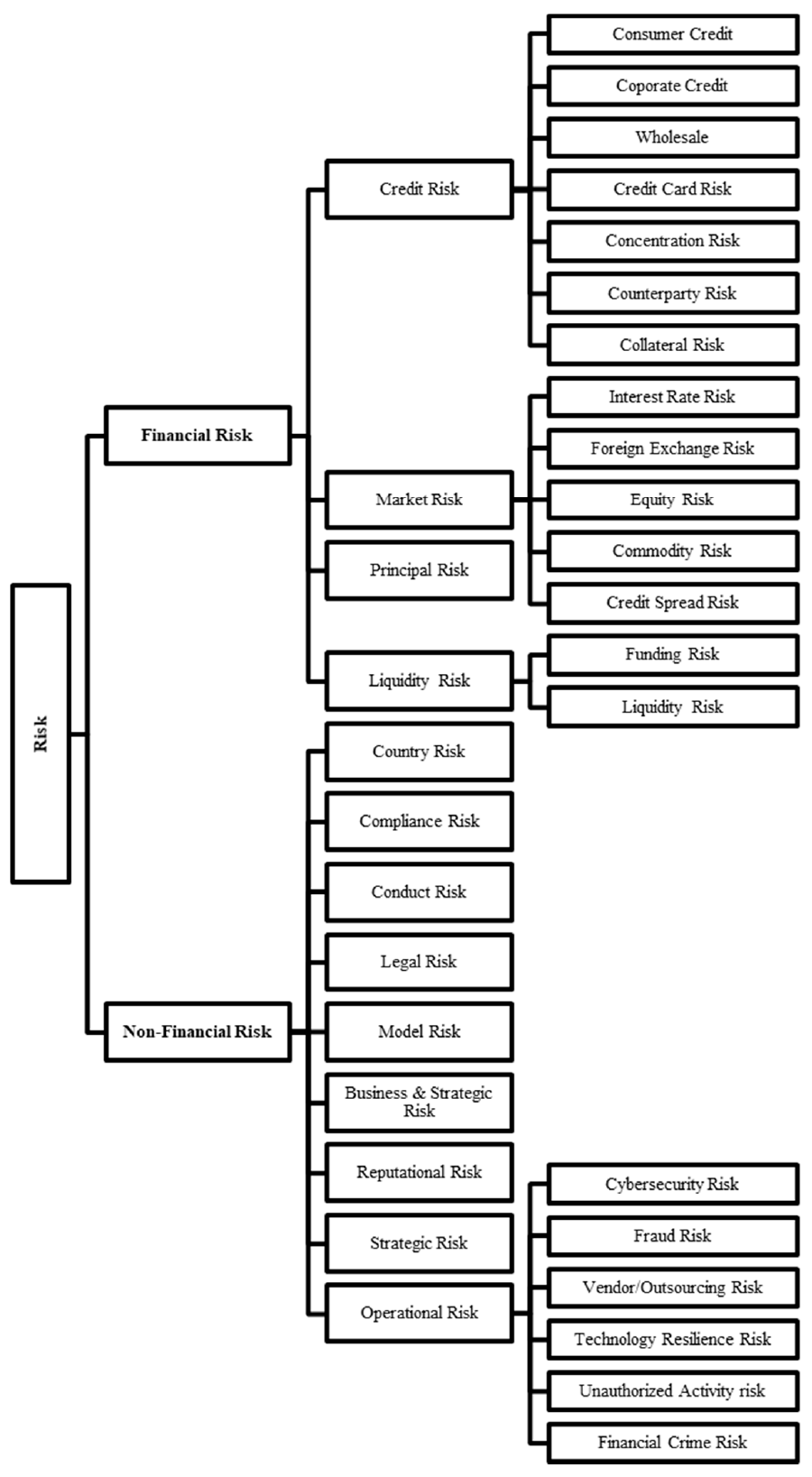}
    \caption{Risk Taxonomy~\cite{leo2019machine}.}
    \label{fig:riskcat}
\end{figure}

\subsubsection{Enrichment Phase}

At this phase, we build domain specific Knowledge Base (KB) based on the risk taxonomy illustrated in Figure~\ref{fig:riskcat}. This risk taxonomy presents most of risk factors related to a company. The branches shows most known risk types such as operational risk, enterprise risk, credit risk, market risk and legal risk factors that have potential to undermine the success of the company. For instance, inside trading could cause the share price of the company to drop. To create a effective banking knowledge base, the risk taxonomy is used as a guide to find and utilize the important databases.
In section~4, the section~4.3.2 explains the design and construction of the risk domain knowledge in banking by a real world example and provide an insight to it. Figure~\ref{fig:DK} illustrates a sample fragment of the risk domain knowledge in banking.


%
We develop a set of enrichment functions to tag the extracted features with their relevancy to the different categories if risk, illustrated in Figure~\ref{fig:riskcat}. For example, if a transaction description contains a keyword or named entity related to the: Sanctioned Destinations (e.g., countries or organizations), or gambling organizations or locations, then the transaction can be tagged with Country, Legal, or Fraud Risk labels.

\begin{algorithm}
 \KwData{Business-Artifact, Domain Knowledge}
 \KwResult{Contextualized Business-Artifact}
 MR Job1: Retrieving Business-Artifact Schema and Preprocessing the artifact;\\
 MR Job2: Extraction:
 \For{Each Business-Artifact} {
  Extract-Feature(keyword, phrase, topic, named entity, sentiment, time, location, etc.);
 }
 MR Job3: Enrichment:
 \For{Each extracted feature} {
  Compute-Similarity(feature,items in the Domain Knowledge);\\
  Link the feature to the similar item in the domain knowledge; 
 }
 MR Job4: Annotation:
 \For{each enriched feature} {
 MTask = Construct the Micro-Task;\\
 Send MTask to the crowd;\\
 Annotate the Item based on crowd Feedback;\\
 Update Training-Data;
 }
 MR Job5: Classification:
 \For{each Business Artifact} {
  Classify the artifact based on the taxonomy of risk in Figure~\ref{fig:riskcat};
 }
 \caption{Banking Data Curation Algorithm.}
\end{algorithm}

\subsubsection{Linking and Analysis Phase}

At this stage, a previous work~\cite{DataSynapse} is leveraged to compute the similarity between an enriched feature and the items exist in the domain knowledge.
This approach leverages cross-document coreference resolution~\cite{CDCR} and intelligent summarization~\cite{SamiSummary} (based on dynamic feature space mapping) to identify the coreferent entities between extracted/enriched items from the business artifacts and the ones that exist in the domain knowledge base.
We develop a scalable algorithm (Algorithm~1) to automate the banking data curation.
This algorithm includes a set of MapReduce Tasks to handle the preprocessing, extraction, enrichment, annotation and risk-based classification of the business artifacts in banking. Since the size of financial datasets are quite large and running statistical or machine learning techniques would consume a long time to process, MapReduce is utilized to distribute the files for processing. Algorithm 1 creates five jobs to execute data curation. The first job retrieves the business-Artifacts schema and prepares it for the high-level feature extraction. The second job extracts the features from the text using several techniques. the next job enriches the extracted information and computes the similarity of them with the domain knowledge. Then, Annotation job sends the result to the domain experts for the feedback. The last job manages the classification for each business artifact.   

\begin{figure}[hp]
    \centering
    \includegraphics[width=1.2\textwidth, angle=90]{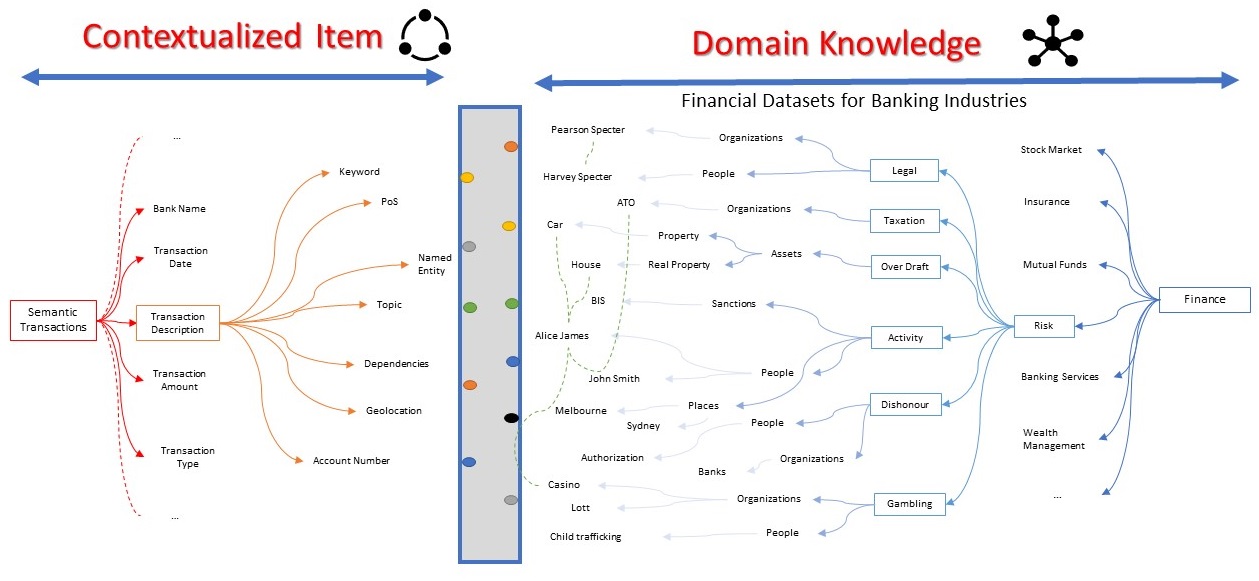}
    \caption{An overview of linking extracted information to the risk-based banking domain.}
    \label{fig:linking}
\end{figure}

Figure~\ref{fig:linking} illustrates the process of connecting business artifacts to the banking domain knowledge. The left side of the figure shows how the extracted information such as keywords, Named Entities and Geo-locations are taken from semantic transactions. On the other hand, it shows the data sources available that could provide details on the risk appetite of the customers. The rectangular indicates the linking stage of the proposed system based on the similarity, matching and classification of the contextualized-item. By utilizing the CDCR-Similarity method, once the semantic item is linked to the entities in the domain knowledge, a relationship will be developed between the two entities that would score the similarity between them. 

\subsubsection{Annotation and Validation Phases}

To determine the risks specific to business artifacts, as an alternate to leveraging the existing measures,
we generate two crowdsourcing micro-tasks and send to domain experts in banking.
The micro-tasks contain the: (i)~actual artifact (e.g., the bank transaction); and (ii)~the contextalized artifact which has been also tagged with the risk-related labels.
The first question in the micro-task is that if the artifact has been correctly identified as a risk-related artifact or not.
The second micro-task is build for the correctly identified risk-related artifacts, and request the knowledge workers to assign a number between 1 to 5 (where 5 is the highest level of risk and 1 is the lowest level of risk) to the business artifact.
As the next step, we develop algorithms leverages this knowledge (provided by the domain experts) to score the risk parameters in the domain specific KB (Figure~\ref{fig:DK}).
As the future and ongoing work, we are developing a Reinforcement learning (RL) algorithm to benefit from the feedback loop (Figure~\ref{fig:framework}) and to
take actions in the banking environment in order to maximize the notion of cumulative reward, and provide more accurate predictions in the construction of the micro-services.

\subsubsection{Summary}

In this section, we focused on understanding the big data in banking sector, provide automated techniques to curate the raw banking data, leveraged the knowledge of banking expert to annotate the curated banking data, and related this curate-annotated banking data to the process analysis in banking.
We proposed an intelligent data-driven pipeline composed of a set of processing elements to move customers' data from one system to another, transforming the banking data into the contextualized data and knowledge along the way.

In the next section, we present an experiment and evaluation of the proposed approach. We present a motivating scenario and develop an intelligent risk-based customer segmentation model to use the banking domain knowledge to extract features from raw data via data mining and crowdsourcing techniques, in the banking domain. 
\section{Experiment and Evaluation}

This section aims to present and evaluate the results obtained from the proposed method and discusses how this method could help the finance industry in terms of risk analysis. The first part of the section describes detailed information on the data input of the system. Source and size of data will be explained how it was obtained to ensure the quality of input is reliable and sensible. Then, the system setup that was used to perform all the experiments will be described to the readers. It would indicate the execution time of the process and the efficiency of the workload.

Moreover, a motivating scenario is presented to familiarise the reader with the financial business process and the life cycle of the proposed method. This case study will illustrate the whole process of contextualising the input data and displaying the output to the analyst in the user interface. A complete discussion will be conducted on the results obtained from the proposed method, and various techniques will be undertaken to evaluate the performance of the system.

\subsection{Dataset}

A dataset has been acquired from Prospa Group Ltd (prospa.com/) which is a private financial institution based in Sydney, Australia . The dataset contains historical data of individual small businesses that are interested in applying for a small business loan with the source of the dataset. The dataset provides a bank transaction of 8483 customers, for the period of last six month before the application date. For privacy reasons, all information on the customers is being replaced by random numbers in the range of 1 to 8483. When a customer authorises the access to their bank statement, variety of information is received such as the method of transaction, device's information involved in the transactions, location of the payer, receipt number, BSB, account number and method of notification.

For instance, when John transfers an amount of \$500 to Monica, the information provided in the dataset could have the following details on the transactions. 

\begin{itemize}
    \item Method of transaction, it shows if the payment was a direct debit, BPay or any other method used to transfer the money. 
    \item  If an online transaction has occurred, information about the type of device used to transfer the fund such as phone number, what type of phone, IP address, name of internet provider and others information related to device and method of connection. 
    \item Geo-location of the customer when transaction happened. 
    \item Account details of the payer and the payee. 
    \item How the payee was notified about the transaction such as SMS, email, or social platforms. 
    \item User information of the payee such as name, address and contact details.
\end{itemize}

However, the source of the database has only provided six columns for this research.  The approved columns of the dataset are as below: 

\begin{itemize}
    \item Customer id: The newly generated integer numbers by the author to protect the identity of the customers. 
    \item Bank name: The name of the bank which the account has been registered. 
    \item Transaction date: The specific day that transaction has taken place shown in the DateTime format.
    \item Transaction type: A categorical data that indicates the transaction that has either entered the account or left the account. 
    \item Transaction description: A text that describes the transaction purposes
    \item Transaction amount: A float data that indicates the amount of money that has been added or removed from the account. 
\end{itemize}

Table \ref{table Dataset} shows a sample of the top three rows of the dataset. As it can be seen six information regarding the transaction could be extracted from the dataset.

\begin{table}[!h]
\caption{ An example of columns of dataset}\label{table Dataset}
\centering
 \begin{tabular}{||c c c c c c||} 
 \hline
 CustomerID & BankName & TransactionDate & TransactionType & TransactionDescription & TransactionAmount \\ [0.5ex] 
 \hline\hline
 1 & ANZ & 2019-06-09 & Credit & EFTPOS TRANSACTION..  & 279.95 \\ 
 \hline
 2 & NAB & 2019-12-12 & Debit & VISA DEBIT .. & -74.05 \\ 
 \hline
 2 & NAB & 2020-07-23 & Debit & TRANSFER DEBIT .. & -190 \\ 
 \hline
 \hline
\end{tabular}
\end{table}

Notably, the applicants and financial institution are based in Australia and the amounts presented in the transaction amount column are in Australian Dollar. Dataset is in CSV format and has a size of 3 GB. Provided the volume, variety and velocity of data, it can be stated that input falls in the concept of Big Data. 

\subsection{System Setup}
Azure Databricks platform is utilized by the author to perform all the experiments of the proposed method. There are three main reasons for using Azure Databricks to perform the operations, which are Running the data on Apache Spark, utilization of Azure Active Directory (AAD) to provide security and easy to collaborate with the data engineering team. According to Lei Gu and Huan Li research on performance evaluation, Apache Spark allows the user to perform the experiments almost 100 times faster than Hadoop and MapReduce technique~\cite{sparkvshadoop}. The configuration of the clusters created on Databricks is presented in table\ref{table SystemSetup}. 

\begin{table}[!h]
\caption{Configuration of the created cluster on Databricks }\label{table SystemSetup}
\centering
 \begin{tabular}{||c c ||} 
 \hline
 Databricks Runtime Version  & 6.3 ML(includes Apache Spark 2.4.4) \\
 \hline
 Worker type & 32.0 GB Memory, 8 cores, 1.5 DBU \\ 
 \hline
 Driver Type & 128.0 GB Memory, 32 Cores, 6DBU \\ 
 \hline
\end{tabular}
\end{table}
A cluster with the above configuration would help to increase the effectiveness of the experiments and reduce the execution time. Moreover, all the codes are developed in Python 3.7.4 for testing purposes and then converted to PySpark 2.4.5 for utilisation on Databricks.

\subsection{Motivating Scenario}

To understand the risk of a process in the money lending industry, history and activities of the borrower are examined to ensure their behaviour is within the risk matrix of the lender. Variety of sources of data are used to search through the history of the borrower and ensuring the money will be used for the right purposes. Aspects such as credit score, occupation, age, income, and other financial commitment have a direct effect on the risk matrix of the application. To enquire and analysis all the requirements of the risk matrix and data manually, the process will take an extended period. Additionally, it would be challenging to conduct an accurate evaluation of the applicant's behaviour and history. An instance of this in the financial sector is when an applicant is an immigrant, and enquiring information about them would require external sources and validation. 

Considering the risk analysis in the financial sector, several bases could have an impact on the determination of policies and implementation of risk matrices. The risk matrix of an application could determine and classify customer behaviour. Several categories could be extracted from a customer's bank statements such as customer risk rating, client type, account type and transaction type. The main known categories that are considered in this project and highlights the risk of the customer are having a debt to taxation office, a dishonour transaction, gambling, mutual transactions with people or organisations under sanctions and infringement. The main challenge would be to provide a clear and real-time insight in terms of identifying the risk rating of the customer. A solution would be to link and classify the customer's transactions to the risk categories and provide an insight to ease up the decision making process. 

This section will focus on a motivating scenario, where an applicant has submitted their bank statement and has requested for a small business loan. The type of information received from the applicant is explained in section 3.1. Efficient analysis of this scenario or any other similar scenario will be discussed in this paper by utilising a three-step data curation pipeline. 

\subsubsection{Step 1:Creating the Semantic Item}

Text mining techniques are required to extract information from a transaction description of a bank statement. There are two types of text mining methods. The first step is to use a traditional low-level method such as cleaning, merging, matching and classifying. The second method would be applying high-level techniques such as Parts of Speech, Named Entities, Dependencies, Stem, Synonyms, Geo-locations and n-gram, which would be able to extract customised keywords and phrases. Combination of these methods would provide more insightful and relevant keywords to the risk analysis. 

Important to realise, the extracted keywords would be able to provide the date, name of the organisation, person and account details. For example, a description such as \emph{WITHDRAWAL AT HANDYBANK AUBURN 2 23965109 22/09/20} indicates that a certain amount of cash has been withdrawn from an ATM number 23965109 at Auburn on 22/09/20 by the cardholder. The first step would be to utilise low-level techniques and feature extraction methods to get some insight into the text. By using these methods, unnecessary punctuation and white spaces will be removed, and all the words will be split into tokens. Then by applying high-level techniques, the system would be able to extract Handybank organization, Auburn suburb, a specific eight-digit number and the date. The Figure below shows the process of creating semantic items from bank transactions.

\begin{figure}[hp]
    \centering
    \includegraphics[width=1.35\textwidth, angle=90]{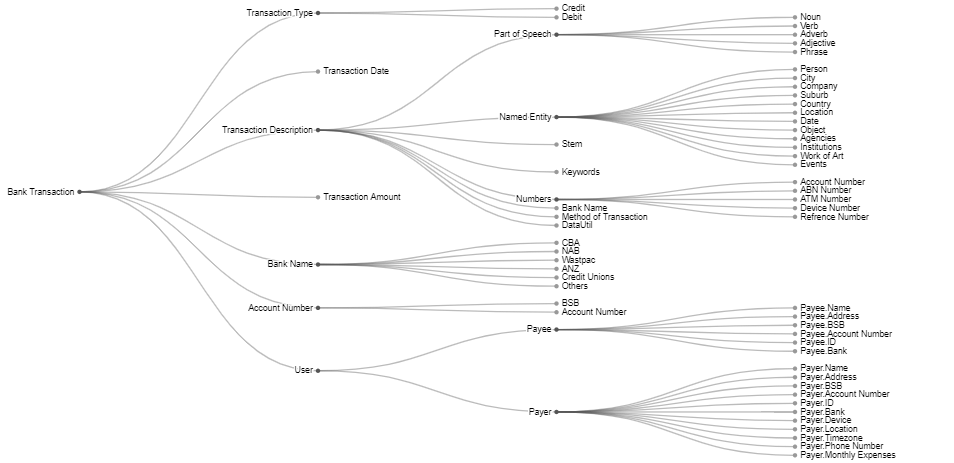}
    \caption{Features and attributes extracted from a single transaction.}
    \label{fig:Semantic Item}
\end{figure}

\textbf{Use Case 1: Extraction on bank statements.} Bank statements are considered as complex and semi-structured data, which would make the analysis more challenging. Let's take Alice James as a small business owner who would like to apply for a small business loan from a financial institution. Alice authorises the institution to get online access to her bank statements for her application to be processed. Alice has provided all the transactions from her business and personal accounts for the past six months. By using text mining algorithms and NLP(Natural Language Programming) techniques, the system can detect the suburbs that Alice visit more often, the name of organisations that she has a connection, all transactions with specific industries and any purchases or transfer of money to outside Australia. By running all the high-level techniques through her bank statement, the system would be able to provide attributes and features from the text of the transaction. Figure \ref{fig:Semantic Item} illustrates all the attributes and features extracted from the bank transaction. 

\subsubsection{Step 2 and 3:Building the Knowledge domain and linking it}
The risk appetite of business is crucially important in the money lending industry. Accordingly, it is critical to develop a risk analysis model and run customer information through them. he model requires a domain knowledge database which could be used to identify the transactions that are related to risk factors. This section will describe the external sources and available databases related to the financial sector. These databases are specific to understanding risk factors of money lending process.

\begin{figure}[hp]
    \centering
    \includegraphics[width=1.25\textwidth, angle=90]{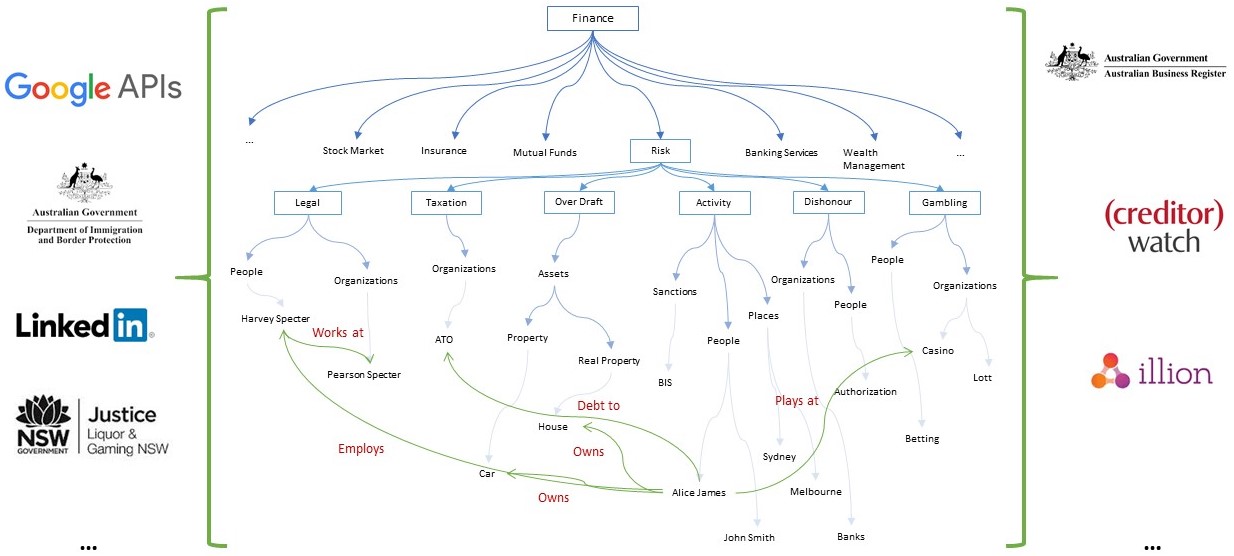}
    \caption{An example of fragment of risk domain knowledge.}
    \label{fig:DK}
\end{figure}

\textbf{Use Case 2: Building Domain Knowledge.} Creating a domain knowledge for the risk appetite in the financial sector requires a variety of related data sources. The first step is identifying the commercial data sources and then finding their relation to \emph{Risk} category. Thinking of Alice case, her information will run through the knowledge domain, and all the related data to her features and attributes will be highlighted. Figure \ref{fig:DK} presents an example of running the Alice case into our domain knowledge and the information extracted from it. On the sides of the brackets, a sample of utilized data services is provided. Around six categories of data are used to construct the background of domain knowledge which are as below\footnote{These data sources are independent from Prospa sources}. 

\begin{itemize}
    \item Legal such as hiring private lawyers, contracting law firms for the small businesses, Freelance lawyers 
    \item Taxation such as any BAS payments, garnishee payments to ATO (www.ato.gov.au/) and not paying taxes.   
    \item Overdraft, meaning to allow the customer to withdraw money from the account with insufficient fund up to the value of their registered assets. 
    \item Conducting business activities with certain people or countries that might be under sanctions. For example having transactions with countries such as Iran, North Korea and Yemen. 
    \item Having dishonour fees, which indicates not having sufficient funding to pay an organization or a person. 
    \item Gambling activities like betting, lottery tickets and casinos.
\end{itemize}

To extract the required data from several sources, a master API has been developed that connects all the APIs from the different data service providers and enables web-based injection. The master API consists of APIs from the following sources. 

\begin{itemize}
    \item Australian Business Register (abr.gov.au/) provided by Australian government
    \item Credit history of business from Creditor Watch (creditorwatch.com.au/)
    \item Credit history of individual business owner from Illion (illion.com.au/)
    \item Citizenship statues from VEVO (immi.homeaffairs.gov.au)
    \item Location, contact details and reviews from Google API (console.developers.google.com)
    \item Business information from Linked In (linkedin.com/)
    \item Liqueur and Gaming registration number from Australian Governments (data.gov.au/)
    \item Industry registration number from state governments (data.nsw.gov.au/)
    \item History of existing customers from local database
\end{itemize}

Furthermore, other APIs such as Wiki-data, Google Knowledge Graph and Word-net are used to enrich the semantic items extracted from the dataset.
For example, All the named entities extracted from Alice's bank statements are linked to Wikipedia to examine if Alice has any relation to banned countries or under-sanction organisations. Consider a transaction \emph{PAYMENT TO GETCAPITAL DEBIT LPT-000457859}, which indicates Alice has financial relationships with another financial institution named Get Capital (getcapital.com.au/). The name \emph{GETCAPITAL} will be linked to a Wikipedia page that shows more data on the organization such as sources of the Get Capital, date of establishment and location of the business. 

\textbf{Use Case 3: Linking the domain knowledge to extracted information from the transactions.} Once all the semantic items are extracted from the bank statements and the domain knowledge has been built. The next challenge would be to link these two datasets and create a contextualized item. To achieve this goal, the author applies the CDCR processing method, which is explained in section 2 of this paper. Cross-Document Coreference Resolution (CDCR) provides a link between these items by identifying the coreference between the extracted entities and the existing ones in the domain. CDCR provides such a service by calculating the similarity among the data objects of both sides. A similarity API has been generated to detect the relationship between dates, numbers, string as well as entities by utilizing similarity techniques such as Jaccard, TF-IDF, cosine, city block and euclidean. For instance, it can be identified from the transaction \emph{MACQUARIE UNIVERSITY MACQUARIE UNI NS AUS Card xx3812 Value Date 04/03/2020} that Macquarie University (mq.edu.au/) is related to the educational sector of the financial dataset.

Figure~\ref{fig:linking} illustrates the process of linking extracted items from Alice's bank statement to the risk-based banking domain knowledge.

\subsection{Risk Analysis and Evaluation}

To evaluate the proposed system, several standard metrics have been utilized to ensure the system is successfully implemented. Around 250,000 transactions over 8465 application have been sampled by operating K-fold cross-validation method. This method has been selected due to its capability of over repeating random sub-sampling, and reduced execution time. For fear that classifying all these transactions might consume a long time from the domain experts, all the duplicate transactions are dropped to confirm the domain experts are not re-classifying the same deals. 

On the subject of the effectiveness, the author has utilized precision(\ref{4.2}), Cohen-Kappa(\ref{4.3}), recall(\ref{4.4}), F-measures(\ref{4.5}) and accuracy(\ref{4.1}) measurements. Accuracy measurement determines the number of correctly classified transactions by domain experts divided by the total generated results, and Precision is the number of correctly classified transactions over total number of transactions. Additionally, Recall is the number of correctly classified transactions over the risk related transactions, and F-measure is the harmonic mean of precision and recall. Cohen-Kappa measure the agreement between the domain expert and the proposed system classification into a mutually exclusive categories. The following formulas shows how these measurements are conducted using mathematical expressions. 

 \begin{multicols}{2}
  \begin{equation}\label{4.1}
    Accuracy = \frac{TP + TN}{(TP+TN+FP+FN)}
  \end{equation}\break
  \begin{equation}\label{4.2}
    Precision = \frac{TP}{(TP+FP)}
  \end{equation}
\end{multicols}

 \begin{multicols}{2}
  \begin{equation}\label{4.3}
    Cohen-Kappa = \frac{p-q}{(1-q)}
  \end{equation}\break
  \begin{equation}\label{4.4}
    Recall = \frac{TP}{(TP+FN)}
  \end{equation}
\end{multicols}

 \begin{multicols}{1}
  \begin{equation}\label{4.5}
    F-measure = \frac{2*Precision*Recall}{Precision+Recall}
  \end{equation}\break
\end{multicols}

In the formulas above, TP stands for true positive, TN means true negative, FP indicates the number of false positives and FN shows the the number of false negatives. For Cohen-Kappa, p is the number of mutual classified transactions divided by total number of transactions, and q is the number of correct classification mines the false classification. 

For further performance analysis of the system, the results of the curated data are compared with classic approaches. The classic approach means using standard text cleaning algorithms and using index searching for matching the strings in the text of description with a list of predefined keywords. If the keyword appears in the document, it will be classified as a risky document.In the classic method, the text is normalized to remove all the noises that could block the system from detecting the pattern of text. Then, most common English words are removed from the text to reduce the computational power and time. Next step is to apply Stemming and Lemmatization methods to reduce words to their root form. Moreover, Part-of-Speech method is applied to categorize the words into tags~\cite{onan2016ensemble} However, the proposed method would link the extracted semantic items with the domain knowledge and then classifies the transaction. The table \ref{table comparison} shows the results of the proposed method compared to classic approaches. 

\begin{table}[h]
\caption{The comparison between the proposed method and classic method}\label{table comparison}
\centering
 \begin{tabular}{||c c c ||} 
 \hline
  & Proposed Method  &  Classic Approach \\
 \hline
 Accuracy  & 0.9152 & 0.8324 \\ 
 \hline
 Recall & 0.8236 & 0.7422 \\ 
 \hline
 Precision & 0.8444 & 0.7713 \\ 
 \hline
 Kappa & 0.8475 & 0.7994 \\ 
 \hline
 F-Measures & 0.8340 & 0.7564 \\ 
 \hline
\end{tabular}
\end{table}

As it can be observed from the table \ref{table comparison}, the performance metrics of the system show that the proposed system has the edge over the classic approaches. 

\subsubsection{Scalability Study}

There is also, however, a further point to be considered. Scalability of the systems is defined within the cluster configuration on Databricks. As discussed in section 3.2, the details of the cluster configuration are presented. In Databricks, the execution time has a direct effect on the cost of the system. To reduce the cost, the author has optimized the programme to ensure the execution time is reduced to 3 minutes using the cluster described in table \ref{table SystemSetup}. However, by applying the classical approach the computation time is around 90 second. The competition time has increased due to having latency to access several databases across the globe as well as keyword extraction techniques. 
\section{Conclusion and Future work}

\subsection{Conclusion}

As technology evolves daily, it is crucial to ensure that financial services are updated and embrace the new technologies. One of the main aspects of the financial sector is understanding and analysis on risk. Having an effective and efficient risk metric allows commercial companies such as a bank to have a clear insight into their customers and products. By utilizing technologies such as natural language processing, machine learning, and artificial intelligence allows banks to develop customer segmentation and analytical segmentation models. Identifying informative attributes of the customers such as customer risk rating, client type, transaction type, revenue forecasting, and expenses classifies the customer within the right risk category. Based on the risk classification, the financial companies can determine the kind of services and products that could help the customer as well as the bank. 

In this dissertation, we focused on understanding the big data in banking sector, provide automated techniques to curate the raw banking data, leveraged the knowledge of banking expert to annotate the curated banking data, and related this curate-annotated banking data to the process analysis in banking.
We proposed an intelligent data-driven pipeline composed of a set of processing elements to move customers' data from one system to another, transforming the banking data into the contextualized data and knowledge along the way. We developed an intelligent risk-based customer segmentation model to 
use the banking domain knowledge to extract features from raw data via data mining and crowdsourcing techniques, in the banking domain. 
The proposed system would use high and low-level engineering approaches on transaction description to extract semantic items. Then, all the extracted information will be linked to the domain knowledge, specifically on banking, using developed APIs to create contextualized data. The contextualized data is used to classify the transaction to either risky or not. Next, a domain expert in the field of credit checking would analyze the input dataset and label each transaction to zero or one. Zero is not risky, and one indicates the deal is related to a risk activity.

Subsequently, several machine learning algorithm is used to evaluate the result of the system and compare the outcome of the proposed method (using contextualized data) with a classic approach which is a combination of regular text cleaning and index searching. The results show the proposed method can classify more transactions compared to conventional methods. The system would allow the financial companies to gain a better insight on their customers and their behaviour by detecting risky activities such as gambling, money laundering, having transactions with sanctioned organizations and people, and importantly determining the availability to payback. 

\subsection{Area of future work}

Given the information in this dissertation, several techniques could enhance the performance of the proposed system. Expanding the domain knowledge, adding more level to the classification of the proposed system and utilizing reinforcement learning (RL) to the domain expert feedback loop are the top three techniques suggested as future work. This section provides a plan of how these techniques could affect the performance of the system in the future and apply it to the application within the risk-based customer segmentation in banking. 

\subsubsection{Towards Intelligent Domain Knowledge}

Intelligent Knowledge Lakes~\cite{intelKG} introduced to facilitate linking Artificial Intelligence (AI) and Data Analytics. A line of future work towards Graph-based modelling and analyzing of banking data~\cite{GraphOlap,DREAM,GraphOlap2,batarfi2015large} to facilitate intelligent feature engineering for risk-based customer segmentation in banking, is to enable AI applications to learn from the domain knowledge and use them to automate banking processes.
Moreover, enabling the analysis of cross-cutting aspects~\cite{CrossCutting} in banking processes, e.g., Provenance~\cite{Provenance}, could improve the accuracy of the proposed system.

For example, to increase the number of identified transactions from a bank statement, it requires to expand and include a deeper search into the customer behaviour. Gaining more access to external databases such as FBI criminal database, Global Risk Database and Interpol databases would assist linking the extracted keywords to the right risk metric. These examples would help to analyse and find the pattern between the applicants and the criminal organizations across the globe. It guides the financial institutions to be a responsible lender and follow the Anti-Money Laundering and Counter-Terrorism Financing (AML/CTF) guidelines. Getting access to the databases requires several security clearances and ethic application to be lodged. 

\subsubsection{Multi-class Pattern Classification Using Neural Networks}

Another line of future work would be to focus on the subjective nature of risk levels. For example, we can introduce more levels to the risk matrix such as five-level rating where they could have different effects on the amount of approved loan and the repayment plan~\cite{schiliro2020novel,CognitiveRS}. By applying a weighting scale to each risk category, the system would be able to differentiate between each risk class and update the scorecard of the financial institution accordingly. For instance, having a transaction to a sanctioned organization in Iran has more risk to it compared to purchasing a lottery ticket. The total weight of the transactions would be summed up at for each application and provide a risk factor to the scorecard. Then the scorecard would be able to determine the liability of the customer based on their behaviour.  

\subsubsection{Developing a Reinforcement Learning}

Another under work approach is to develop a reinforcement learning (RL) to apply the feedback loop~\cite{feedback1,feedback2,feedback5} to the domain expert output. It would result in training the system to provide higher accuracy. By having a higher prediction accuracy, real-time decision making could be implemented. This would help the finance sector in delivering faster and more efficient services to the customers. The current proposed method assumes the credit officer are fully reliable and are completely correct. In terms of risk assessment, human error has been identified as one of the factors that could impact the system. To reduce risks caused by the domain experts, it is highly recommended to apply the reinforcement technique as a feedback loop to minimize the risk. 

\section*{Acknowledgements}
- I acknowledge the AI-enabled Processes (AIP\footnote{https://aip-research-center.github.io/}) Research Centre and Prospa\footnote{https://www.prospa.com/} for funding this project.

\bibliographystyle{abbrv}
\bibliography{ms}

\end{document}